\documentclass[compsoc, conference, a4paper, 10pt, times]{IEEEtran}

\usepackage{cite}
\usepackage{amsmath,amssymb,amsfonts}
\usepackage{algorithmic}
\usepackage{graphicx}
\usepackage{textcomp}
\usepackage{xcolor}
\usepackage{booktabs}
\usepackage[hyphens]{url}
\usepackage[hidelinks]{hyperref}

\usepackage[nolist]{acronym}
\providecommand{\Ac}{\ac}
\providecommand{\Acp}{\acp}
\providecommand{\Acf}{\acf}
\acrodefplural{oem}[OEMs]{Original Equipment Manufacturers}
\acrodef{ck}[cK]{component key}
\acrodef{rk}[rK]{root key}
\acrodef{dma}[DMA]{Direct Memory Access}
\acrodef{intelme}[IntelME]{Intel Management Engine}
\acrodefplural{soc}[SoCs]{systems on a chip}
\acrodefplural{tpm}[TPMs]{Trusted Platform Modules}
\acrodef{ftpm}[fTPM]{firmware TPM}
\acrodef{dtpm}[dTPM]{discrete TPM}
\acrodefplural{pcr}[PCRs]{Platform Configuration Registers}
\acrodef{vmk}[VMK]{Volume Master Key}
\acrodef{fvek}[FVEK]{Full Volume Encryption Key}
\acrodef{txt}[TXT]{Trusted Execution Technology}
\acrodef{tocttou}[TOCTTOU]{Time-of-check to time-of-use}
\acrodef{mac}[MAC]{message authentication code}
\acrodef{hmac}[HMAC]{hash-based message authentication code}
\acrodef{kdf}[KDF]{key derivation function}
\acrodef{pcb}[PCB]{printed circuit board}
\acrodef{pc}[PC]{program counter}
\acrodef{asid}[ASID]{Address Space Identifier}
\acrodef{pdh}[PDH]{Platform Diffie-Hellman Key}
\acrodef{rmp}[RMP]{Reverse Mapping}
\acrodef{em}[EM]{electromagnetic}
\acrodef{ma}[MA]{Migration Agent}
\acrodef{oek}[OEK]{Offline Encryption Key}
\acrodef{id}[ID]{256-bit identifier}
\acrodef{psp-os}[PSP OS]{PSP OS}
\acrodef{fde}[FDE]{Full Disk Encryption}

\acrodef{os}[OS]{Operating System}
\acrodef{nv}[NV]{non-volatile}
\acrodef{oem}[OEM]{Original Equipment Manufacturer}
\acrodef{cli}[cli]{Command-Line Interface}
\acrodef{ide}[IDE]{Integrated Development Environment}

\acrodef{soc}[SoC]{System-on-a-Chip}
\acrodefplural{soc}[SoCs]{Systems-on-a-chip}
\acrodef{ic}[IC]{Integrated Circuit}
\acrodef{tee}[TEE]{Trusted Execution Environment}
\acrodef{dut}[DUT]{Device Under Test}
\acrodefplural{dut}[DUTs]{Devices Under Test}

\acrodef{uefi}[UEFI]{Unified Extensible Firmware Interface}
\acrodef{pi}[PI]{Platform Initialization}

\acrodef{sec}[SEC]{Security}
\acrodef{pei}[PEI]{Pre-EFI Initialization}
\acrodef{dxe}[DXE]{Driver Execution Environment}
\acrodef{pk}[PK]{Platform Key}
\acrodef{kek}[KEK]{Key Encryption Key}
\acrodef{pin}[PIN]{Personal Identification Number}

\acrodef{tcg}[TCG]{Trusted Computing Group}
\acrodef{tpm}[TPM]{Trusted Platform Module}
\acrodef{rot}[RoT]{Root of Trust}
\acrodef{tss}[TSS]{Trusted Software Stack}
\acrodef{pcr}[PCR]{Platform Configuration Register}
\acrodef{srtm}[SRTM]{Static Root of Trust Measurement}
\acrodef{vm}[VM]{Virtual Machine}
\acrodef{tcb}[TCB]{Trusted Computing Base}
\acrodef{mitm}[MITM]{Man-in-the-Middle}

\acrodef{sgx}[SGX]{Software Guard Extensions}

\acrodef{vr}[VR]{Voltage Regulator}

\acrodef{fi}[FI]{Fault Injection}
\acrodef{vid}[VID]{Voltage Identification}

\acrodef{csp}[CSP]{Cloud Service Provider}
\acrodef{vpn}[VPN]{Virtual Private Network}
 
\acrodef{ecb}[ECB]{Electronic Codebook}
\acrodef{cbc}[CBC]{Cipher Block Chaining}
\acrodef{ctr}[CTR]{Counter}
\acrodef{iv}[IV]{Initialization Vector}

\acrodef{spi}[SPI]{Serial Peripheral Interface}
\acrodef{lpc}[LPC]{Low Pin Count}
\acrodef{cs}[CS]{Chip Select}
\acrodef{i2c}[I²C]{Inter-Integrated-Circuit}
\acrodef{ack}[ACK]{Acknowledgement}
\acrodef{gpio}[GPIO]{General Purpose Input-Output}

\acrodef{msb}[MSB]{Most Significant Bit}

\acrodef{rom}[ROM]{Read-Only Memory}
\acrodef{bios}[BIOS]{Basic Input/Output System}
\acrodef{mmio}[MMIO]{Memory Mapped Input/Output}

\acrodef{amd}[AMD]{AMD}

\acrodef{amd-sp}[AMD-SP]{AMD Secure Processor}
\acrodef{psp}[PSP]{Platform Security Processor}
\acrodef{ccd}[CCD]{Core Complex Die}
\acrodef{psb}[PSB]{\emph{Platform Secure Boot}}
\acrodef{sev}[SEV]{Secure Encrypted Virtualization}
\acrodef{sev-es}[SEV-ES]{SEV Encrypted State}
\acrodef{sev-snp}[SEV-SNP]{SEV Secure Nested Paging}
\acrodef{smu}[SMU]{System Management Unit}
\acrodef{ikek}[iKEK]{Intermediate Key Encryption Key}
\acrodef{ccp}[CCP]{Crypto Co-Processor}
\acrodef{smn}[SMN]{System Management Network}
\acrodef{lsb}[LSB]{Local Storage Buffer}

\acrodef{svi2}[SVI2]{Serial Voltage Identification Interface 2.0}

\acrodef{vcore}[$V_{Core}$]{$V_{Core}$}
\acrodef{vsoc}[$V_{SoC}$]{$V_{SoC}$}

\acrodef{svd}[SVD]{\ac{svi2} Data}
\acrodef{svc}[SVC]{\ac{svi2} Clock}
\acrodef{svt}[SVT]{\ac{svi2} Telemetry}
\acrodef{pwrok}[PWROK]{Power okay}

\acrodef{psil}[PSI\_L]{Power State Indicator Level}
\acrodef{llst}[LLST]{Load-Line Slope Trim}
\acrodef{tfn}[TFN]{Telemetry Functionality Flag}

\acrodef{votf}[VOTF]{VID-on-the-Fly}

\acrodef{ffs}[FFS]{Firmware File System}

\acrodef{ark}[ARK]{AMD Root Key}
\acrodef{ask}[ASK]{AMD SEV Key}
\acrodef{kds}[KDS]{Key Distribution System}
\acrodef{cek}[CEK]{Chip Endorsement Key}
\acrodef{svn}[SVN]{security version number}
\acrodef{vcek}[VCEK]{Versioned Chip Endorsement Key}
\acrodef{(v)cek}[(V)CEK]{(Versioned) Chip Endorsement Key}

\acrodef{csme}[CSME]{Converged Security and Manageability Engine}
\acrodef{ptt}[PTT]{Platform Trust Technology}

\acused{vcore}
\acused{vsoc}
\acused{(v)cek}

\usepackage{csquotes}
\usepackage[binary-units,abbreviations]{siunitx}
\DeclareSIUnit\Mbps{\mega\bit\per\second}
\usepackage{enumitem}

\begin{document}

\title{faulTPM: Exposing AMD fTPMs' Deepest Secrets}

\author{
\IEEEauthorblockN{Hans Niklas Jacob\IEEEauthorrefmark{1},
Christian Werling\IEEEauthorrefmark{1},
Robert Buhren,
Jean-Pierre Seifert\IEEEauthorrefmark{2}}
\IEEEauthorblockA{
\textit{Technische Universität Berlin -- SecT}\\
\IEEEauthorrefmark{2}{\small also:} \textit{Fraunhofer SIT}\\
\{
  \href{mailto:hnj@sect.tu-berlin.de}{hnj},
  \href{mailto:cwerling@sect.tu-berlin.de}{cwerling},
  \href{mailto:robert.buhren@sect.tu-berlin.de}{robert.buhren},
  \href{mailto:jpseifert@sect.tu-berlin.de}{jpseifert}
\}\href{mailto:hnj@sect.tu-berlin.de,cwerling@sect.tu-berlin.de,robert.buhren@sect.tu-berlin.de,jpseifert@sect.tu-berlin.de}{@sect.tu-berlin.de}
}
}

\maketitle

\makeatletter
\begingroup
\renewcommand\thefootnote{\IEEEauthorrefmark{1}}
\def\@IEEEcompsocmakefnmark{\hbox{\normalfont\@thefnmark\ }}
\footnotetext{Both authors contributed equally.}
\endgroup
\makeatother

\begin{abstract}
\Acp{tpm} constitute an integral building block of modern security features.
Moreover, as Windows 11 made a \ac{tpm} 2.0 mandatory, they are subject to an ever-increasing academic challenge.
While \acp{dtpm} -- as found in higher-end systems -- have been susceptible to attacks on their exposed communication interface, more common \acp{ftpm} are immune to this attack vector as they do not communicate with the CPU via an exposed bus.
\\
    In this paper, we analyze a new class of attacks against fTPMs:
    Attacking their \ac{tee} can lead to a full TPM state compromise.
    We experimentally verify this attack by compromising the \ac{amd-sp}, which constitutes the \ac{tee} for AMD's \acp{ftpm}.
In contrast to previous \ac{dtpm} sniffing attacks, this vulnerability exposes the complete internal \ac{tpm} state of the \ac{ftpm}.
It allows us to extract any cryptographic material stored or sealed by the \ac{ftpm} regardless of authentication mechanisms such as \ac{pcr} validation or passphrases with anti-hammering protection. 
\\
First, we demonstrate the impact of our findings by -- to the best of our knowledge -- enabling the first attack against \ac{fde} solutions backed by an \ac{ftpm}.
    Furthermore, we lay out how any application relying solely on the security properties of the TPM -- like \emph{Bitlocker's TPM-only} protector -- can be defeated by an attacker with 2-3 hours of physical access to the target device.
Lastly, we analyze the impact of our attack on \ac{fde} solutions protected by a \emph{TPM and PIN} strategy.
While a naive implementation also leaves the disk completely unprotected, we find that \emph{BitLocker's} \ac{fde} implementation withholds some protection depending on the complexity of the used PIN.
Our results show that when an \ac{ftpm}'s internal state is compromised, a \emph{TPM and PIN} strategy for \ac{fde} is less secure than TPM-less protection with a reasonable passphrase.

\end{abstract}

\acresetall

\section{Introduction}

\Ac{tpm} is a standard for a dedicated subsystem providing security primitives to a system.
\acp{tpm} include a hardware random number generator, can measure and attest system state, securely generate cryptographic keys, and provide an interface for protecting sensitive data akin to a smartcard.

Initially, \acp{tpm} were exclusively implemented as discrete chips connected to the CPU via buses on the mainboard.
However, CPU vendors introduced \acp{ftpm} as an alternative implementation approach \cite{lai_amd_2013, intel_corporation_whitepaper_2014, raj_ftpm_2016}. 
In contrast to \acp{dtpm}, \acp{ftpm} are implemented through code running in a \ac{tee} provided by the CPU vendor.
While \acp{dtpm} are still found in higher-end devices,
\acp{ftpm}' omnipresence in modern CPUs made them an affordable and convenient alternative for \acsp{oem}.
In addition, the ever-increasing integration of TPM services into cryptographic APIs, e.g., \emph{Platform Crypto Provider} in Microsoft Windows \cite{microsoft_how_2021}, fosters \ac{tpm} usage across many layers and domains.

One of the more prominent use cases for a \ac{tpm} is \acf{fde}, as it can secure \emph{data-at-rest} without requiring the user to memorize a pre-boot passphrase.
The standardized interplay of firmware and TPM, commonly called \emph{Measured Boot}, verifies that the system was booted into a trusted state before \ac{fde} keys are released from the \ac{tpm} to the \ac{os}.
Hereby, the system measures each firmware component and configuration value into the TPM, which keeps a cryptographically secure log of these measurements.
Once the \ac{os} is in a trusted state, as reflected by the TPM's measurement logs, the TPM provides access to the disk encryption keys.
Since the \ac{os} is now already in control of the system and its connected devices, it can effectively employ security measures like a login prompt to enforce access control to the user's data.

\subsubsection*{Threat model}

We assume an attacker with prolonged physical access to a device, e.g., a stolen or lost laptop.
While this puts the data on the laptop at immediate risk, TPM-based \emph{Measured Boot} should be able to protect private data, such as a VPN key or FDE keys, from the attacker.
This means only a properly booted runtime is able to access the sealed data.
For this paper, vulnerabilities in the UEFI firmware or \ac{os} are out of scope.

Scenarios where this threat model becomes relevant are journalists protecting their sources or traveling employees carrying intellectual property on their company-provisioned laptop.

\subsubsection*{AMD's fTPM}

While (d)TPMs were previously targeted at professional users and use cases only, the adoption of \acp{ftpm} highlights that \acp{tpm} of all sorts must undergo security testing. 
Moreso, since Windows 11 introduced a \ac{tpm} 2.0 requirement \cite{microsoft_windows_2021}, which in many devices like consumer-level laptops is available only as an \ac{ftpm}.

In the past, \acp{dtpm} have been susceptible to attacks compromising their communication buses \cite{winter_hijackers_2013, andzakovic_extracting_2019,  dewaele_tpm_2021}.
\acp{ftpm}, however, do not expose their communication with the CPUs, making this particular attack vector infeasible.
Nevertheless, other attack vectors have come to light, especially concerning their execution environment.

While the dominant vendor for CPUs is Intel, AMD’s market share has been at a record high recently \cite{alcorn_amd_2022-1}.
The execution environment for \acp{ftpm} on AMD systems is the \ac{amd-sp}, which has recently been subject to physical and firmware attacks \cite{buhren_insecure_2019,buhren_one_2021}.
In these works, the authors analyzed to what extent the \ac{sev} technology is affected by these security issues. 
However, the authors neglected to analyze the impact of this \ac{rot} compromise on other critical security functionality provided by the \ac{amd-sp}.

In light of these attacks, we ask: What are the practical consequences of physical attacks against AMD's \acp{ftpm}?

\subsection{Our Contributions}

In this paper, we show that \acs{amd}'s \acp{ftpm} are vulnerable to
physical attacks against their execution environment: the \ac{amd-sp}.
Our attack utilizes the \ac{amd-sp}'s vulnerability to \emph{voltage fault injection} attacks \cite{buhren_one_2021} to extract a chip-unique secret from the targeted CPU.
This secret is subsequently used to derive the storage and integrity keys protecting the \ac{ftpm}'s \ac{nv} data stored on the \ac{bios} flash chip.

In contrast to the previous \ac{dtpm} sniffing attacks, our approach exposes the complete internal \ac{tpm} state of the \ac{ftpm}.
This allows us to extract any cryptographic material stored or sealed by the \ac{ftpm} regardless of the authentication mechanisms, such as the measured system state
or passphrases with anti-hammering protection.
Additionally, once we have extracted the chip-unique secret through \emph{voltage fault injection}, we only need to re-read the \ac{bios} flash to defeat any \ac{tpm}-based security measures.

To demonstrate the impact of our findings, we re-enable attacks against \ac{fde} solutions that rely solely on the \ac{tpm}.
Furthermore, we analyze how our findings impact an \ac{fde} key protected by a \emph{TPM and PIN} strategy.
We find that Microsoft \emph{BitLocker's} \ac{fde} implementation is reduced to the security of a (TPM-less) \emph{PIN-only} strategy.
In contrast, a naive implementation leaves the disk completely unprotected once the \ac{tpm} is defeated.
The example of \emph{systemd-cryptenroll}, a tool for enrolling hardware security tokens and devices into a LUKS2 encrypted volume, demonstrates this potential weakness.

In summary, our contributions are:
\begin{itemize}
    \item We reverse-engineer the \ac{nv} storage format of AMD's \ac{ftpm} and the derivation of the chip-unique keys protecting its confidentiality and integrity.
    \item We leverage previously published hardware vulnerabilities on the \ac{amd-sp} to extract the cryptographic seeds used to derive the \ac{nv} storage keys.
    \item Using the decrypted NV storage, we can extract any cryptographic secret and unseal arbitrary TPM objects protected with the \ac{ftpm}.
    \item We use this ability to successfully attack Microsoft BitLocker's \emph{TPM-only} key protector.
    \item We analyze the security of \emph{TPM and PIN} protectors for \ac{fde} keys and describe how BitLocker withstands a compromised TPM when a strong PIN is used while a naive implementation does not.
    \item  We publish all required tools to mount the attack in \cite{jacob_pspreverseftpm_attack_2023}.
\end{itemize}

All security-relevant findings discussed in this paper were responsibly disclosed to AMD, Microsoft, and the systemd-cryptenroll maintainers.
The systemd-cryptenroll maintainers quickly got back to us to discuss specific mitigation strategies.

\section{Background}
\label{sec:background}

In this Section, we will introduce basic concepts necessary for our paper, i.e., \aclp{tpm}, \acl{fde} in general, Microsoft BitLocker in particular, and the \acl{amd-sp}.

\subsection{Trusted Platform Modules}
\label{sec:background:tpm}

\Acp{tpm} are secure cryptoprocessors specified by the \acf{tcg} featuring a hardware random number generator, secure generation of cryptographic keys, \acfp{pcr} to measure a system's boot process, and secure storage/sealing of those keys or user-provided data.
The `TPM 2.0 Library specification' \cite{trusted_computing_group_trusted_2019} (also known as ISO/IEC 11889) specifies how a \ac{tpm} should interact with an external system.
Two distinct types of \acp{tpm} can be distinguished based on their implementation: \acfp{dtpm} and \acfp{ftpm} \cite{trusted_computing_group_tpm_2019}. 
The implementation type has different implications on the security level of the TPM. 
However, regardless of whether a \ac{dtpm}, or a \ac{ftpm} is used, the provided functionality is the same. 
\\
Both the security implications of the two implementation approaches, as well as the basic TPM functionality, are explained in the following sections:

\subsubsection{Discrete TPM}
\label{sec:background:tpm:discrete}

\acp{dtpm} are dedicated hardware components implementing the \ac{tpm} specification and are commonly built into commercial business laptops.
\acp{dtpm} implement a degree of physical tamper resistance to protect the stored secrets from exposure and are therefore considered the most secure \ac{tpm} variant \cite{trusted_computing_group_tpm_2019}.
Nonetheless, passive physical attacks have been demonstrated that target the communication channel between the \ac{dtpm} and the rest of the system \cite{winter_hijackers_2013}.
These can be used to, e.g., circumvent \ac{tpm}-based \ac{fde} solutions that unlock a disk without the need for a passphrase or PIN.

\subsubsection{Firmware TPM}
\label{sec:background:tpm:firmware}

\acp{ftpm}, on the other hand, implement \ac{tpm} functionality mainly through software running in a \acf{tee} available to the CPU.
Intel and AMD provide \acp{ftpm} with their Desktop CPUs, running on the Intel Management Engine (Intel-ME) and \ac{amd-sp}, respectively.
Due to the integration of these coprocessors into the main die, they do not require external buses to communicate with the CPU.
Therefore, sniffing attacks similar to the \ac{dtpm} attacks would require costly on-die probing, which -- to the best of our knowledge -- has not been shown against \acp{ftpm} yet.
On the other hand, \acp{ftpm} rely heavily on the security properties of the \ac{tee}.

\subsubsection{Sealing}
\label{subsubsec:sealing}

A \ac{tpm} can \emph{seal} data, such as cryptographic keys or secrets, and regulate access to the data.
For example, a user might generate an RSA key on the \ac{tpm}, protect it with a passphrase, and limit it to be used as a signing key that can not be exported from the \ac{tpm}, \emph{binding} the key to the device.
In such a scenario, the \ac{tpm} acts like a smartcard.

Since \acp{tpm} have only a limited amount of \acf{nv} storage on the device itself, data is usually stored externally in a \emph{sealed object} and can be loaded into the \ac{tpm} when needed.
While these external objects may reveal details about the kind of object that is sealed, the sensitive data, e.g., RSA private key, is encrypted and signed with storage keys only accessible to the \ac{tpm}.
The format of these external objects and the key derivation for their protecting keys are specified in the `TPM 2.0 Library Specification' \cite{trusted_computing_group_trusted_2019}.\footnote{
	See `22 Protected Storage' of `Part 1: Architecture' for the key derivation algorithms and, for the object format, \texttt{TPM2B\_PUBLIC}/\texttt{TPM2B\_PRIVATE} in `Part 2: Structures'.
}

The authorization options for sealed \ac{tpm} objects are manifold and can be combined with `and' and `or' clauses to form complex authorization policies \cite{arthur_practical_2015}.
Two policies commonly used are \ac{pcr} authorization and \acs{pin}/passphrase authorization with anti-hammering protections.

\subsubsection{Platform Configuration Registers}
\label{sec:background:tpm:pcr}

A \ac{pcr} is a memory location in the \ac{tpm} used to store a hash, e.g., a SHA256 digest.
\Acp{pcr} can only be updated by extending the existing value with a new value as follows \cite{arthur_practical_2015}:
$$ PCR[n] = hash\_alg( \; PCR[n] \; || \; ExtensionValue \; ) $$
The only way to reset a \ac{pcr} is to reset the \ac{tpm}, in which case it will reset to zero.
This means that during a system's uptime, the \ac{pcr} acts as a secure record of all extension values:
Once a value has been recorded, one cannot 'un-record' it, i.e., force the \ac{pcr}'s value back to a previous state -- unless the underlying cryptographic hash algorithm is flawed.

In the \emph{measured boot} scenario -- one of the most prominent use-cases for \acp{pcr} -- each firmware component involved during boot is hashed and extended into an appropriate \ac{pcr} \emph{before} it is executed.
In order to protect against malicious code running early in the boot process, an operating system can record a set of known-secure \ac{pcr} values and verify these \acp{pcr} on each boot.
For example, on a PC platform, \acp{pcr} zero to seven record the system's boot process.

The \ac{tpm} can further use the \acp{pcr} itself to authorize access to a \ac{tpm} object.
To check a \ac{pcr} authorization policy, the \ac{tpm} compares the current \acp{pcr} to a set of known values included with the policy.
Such a policy can, e.g., protect a cryptographic key from  being used when the system has been infected with a \emph{Root- or Bootkit}.

\subsubsection{PINs, Passphrases, and Anti-Hammering}
\label{sec:background:tpm:pins}

Another method of authorizing access to sealed \ac{tpm} objects is a passphrase or \ac{pin} \cite{arthur_practical_2015}.
Of course, passphrase authorization is also commonly used for software-only protection, e.g., with SSH keys.
However, \ac{tpm}-based passphrase authorization can defend much more effectively against dictionary or brute-force attacks.
The \ac{tpm} can be configured to keep track of the failed passphrase authorization attempts and limit the amount or rate of authorization attempts.
For example, Windows limits the authorization rate to one try every ten minutes once 32 failed attempts have been made \cite{microsoft_bitlocker_2021-1}.
This significantly limits the power of dictionary or brute-force attacks and even enables the use of lower entropy \acp{pin} to protect a \ac{tpm} object.

\subsubsection{Applications}
\label{sec:background:tpm:applications}

A significant application of \acp{tpm} is hardware-aided key management.
The \emph{tpm2-pkcs11} project \cite{tpm2-software_tpm2-pkcs11_2022}, for example, exposes the \ac{tpm}'s functionality as a \emph{PKCS \#11} interface, which is a standardized API to access cryptographic services like smartcards \cite{gleeson_pkcs_2015}.
With \emph{tpm2-pkcs11} the \ac{tpm} can be used to, e.g., hold and manage an SSH key protected with anti-hammering and bound to the system's \ac{tpm}.
Similarly, Microsoft uses the \ac{tpm} as a hardware backend for their \emph{Platform Crypto Provider} API in Windows \cite{microsoft_how_2021}.
Additionally, \emph{Windows Hello} uses a device-unique asymmetric key-pair protected to authenticate a device with the identity provider (Windows).
This key is sealed by and bound to the \ac{tpm} of the device, as well as protected by \ac{pin} that authenticates the user towards the device \cite{microsoft_how_2021}.
Another prominent use case for \acp{tpm} is \acf{fde}, which we discuss at length in the next Section.

\subsection{Full Disk Encryption}
\label{sec:background:fde}

\Acf{fde} protects the confidentiality and integrity of \emph{data-at-rest} of a computer, i.e., the contents of the computer's disks.
This includes the computer's operating system, which means that the disk's encryption keys need to be available to the boot loader before it can load the operating system.
One approach asks the user to enter a passphrase in a pre-boot environment and derive a key from this passphrase.
However, to support changing the passphrase and allow multiple decryption methods, \ac{fde} tools like \emph{BitLocker} or \emph{LUKS} do not directly seal the key encrypting the \emph{data-at-rest}, but store multiple encrypted copies of this key alongside the data \cite{microsoft_manage-bde_2022, broz_luks2_2022}.
The respective \emph{key-encryption keys} for these encrypted copies represent a method of decrypting the disk, e.g., a recovery key stored safely out-of-band, a passphrase, or a \ac{tpm}-based decryption method.
For a \ac{tpm}-based decryption method, the \ac{fde} tools seal the \emph{key-encryption key} with the \ac{tpm} and protect it using the \ac{tpm}'s authentication mechanisms.

\subsubsection{TPM-only strategy}
\label{sec:background:fde:tpm_only}

Here, the sealed \emph{key-encryption key} is protected solely by a \ac{pcr} policy (\ref{sec:background:tpm:pcr}).
The \ac{pcr} values necessary to unseal the key reflect a boot with a trusted firmware and boot loader configuration.
If any part of the boot process is altered, e.g., by a UEFI Root- or BootKit, the \ac{pcr} values reveal this change to the \ac{tpm}, and the (possibly compromised) boot loader cannot unseal the \emph{key-encryption key}.
On the other hand, if the key can be unsealed -- meaning the \ac{pcr} values indicate a boot with valid and trusted firmware -- the operating system's security measures enforce access policies to the disk's content and \emph{data in use}, including the unsealed key.

A notable distinction of this approach is its transparency toward the user.
Although no interaction is required in the pre-boot environment, even an attacker with physical access cannot access the \emph{data-at-rest} off the disks.
However, a downside of \emph{TPM-only} strategies for \ac{fde} is that they offer no additional security if the \ac{tpm} protection can be broken, e.g., by a sniffing attack (\ref{sec:related_work:bitlocker_attacks}).

\subsubsection{TPM and PIN strategy}
\label{sec:background:fde:tpm_and_pin}

A strategy that offers more protection than the \emph{TPM-only} strategy against, e.g., sniffing attacks, is to include a PIN or passphrase.
The \ac{tpm}'s anti-hammering features (\ref{sec:background:tpm:pins}) compensate even the use of a lower-entropy PIN compared to a (TPM-less) \emph{passphrase-only} strategy that can only protect against brute-forcing through a strong secret.
If, on the other hand, an attacker gains access to the PIN or passphrase, the \ac{pcr} policy still ensures that the key can only be accessed by the trusted boot loader running alongside the trusted firmware, leaving other (OS-level) protection mechanisms intact.

\subsection{Microsoft BitLocker}
\label{sec:background:bitlocker}

BitLocker is a full-volume encryption feature by Microsoft integrated into Windows and available since Windows Vista. 
BitLocker aims to protect the confidentiality and integrity of \emph{data-at-rest}, i.e., when the computer is powered off or in hibernate, from unauthorized access.

\begin{displayquote}
    BitLocker Drive Encryption is a data protection feature that integrates with the operating system and addresses the threats of data theft or exposure from lost, stolen, or inappropriately decommissioned computers.
    BitLocker provides the most protection when used with a Trusted Platform Module [...].
    \cite{microsoft_bitlocker_2021-2}
\end{displayquote}

On Windows 11, BitLocker is enabled by default on systems with an enabled \ac{tpm} and when a Microsoft Account is used during setup \cite{yip_bitlocker_2022}. 
\\
In the Windows runtime, BitLocker's deep integration into Windows makes the encryption and decryption of data completely transparent to user applications.
During boot, BitLocker will provide an early (unencrypted) boot component handling the decryption of the remaining disk.

\subsubsection{Key management}

\begin{figure}[h]
\includegraphics[width=\columnwidth]{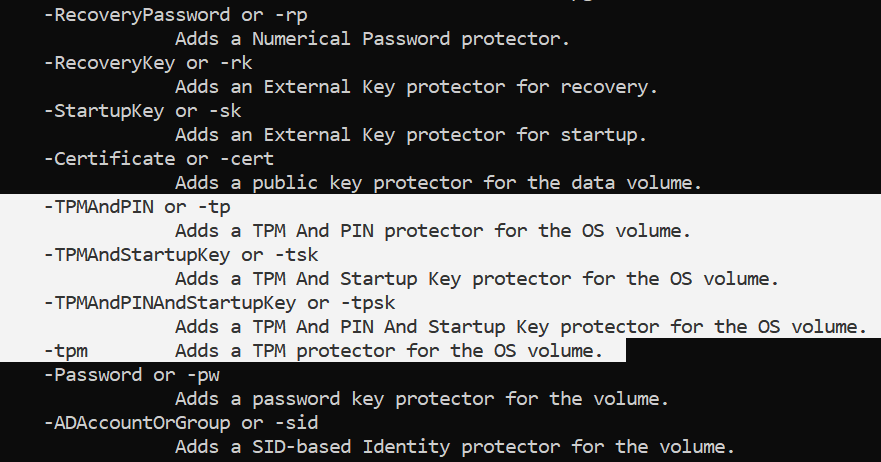}
\caption{Available protector types for Microsoft BitLocker on Windows 11, as displayed by `manage-bde'. \ac{tpm}-related protectors are highlighted.}
\label{fig:available_protectors}
\end{figure}

All data on disk is encrypted using the Full Volume Encryption Key (FVEK).
The FVEK is stored encrypted by the \ac{vmk} in an unencrypted portion of the volume.
The \ac{vmk} is stored encrypted by several so-called \emph{(key) protectors} that can be seen in Figure \ref{fig:available_protectors}, one of which is, by default, the numeric recovery key a user is requested to print out during setup.
The encrypted copies of the \ac{vmk} -- by default, a \ac{tpm}-protected \ac{vmk} and a recovery key-protected \ac{vmk} -- are stored in an unencrypted portion of the BitLocker-protected volume.
They can be managed on Windows through a Microsoft-provided PowerShell tool (`manage-bde`) and Linux with the third-party tool \emph{Dislocker} \cite{aorimn_dislocker_2022}.
\emph{Dislocker} is furthermore able to mount a BitLocker-encrypted volume, given an unencrypted \ac{vmk}.

\subsubsection{TPM-based protectors}

Besides protector modes like a pass\-phrase or USB key, BitLocker provides multiple protectors facilitating a \ac{tpm}.
In fact, BitLocker, by default, uses the \ac{tpm}'s boot integrity measurements exclusively to ensure that only an untampered BitLocker runtime can access the \ac{tpm}-protected \ac{vmk}.
\\
During setup, BitLocker takes the \ac{vmk} and \emph{seals} it with the current \ac{tpm} state represented by the PCR register values 0, 2, 4, and 11.
It then stores a handle to this \emph{sealed} \ac{tpm} object in the volume's unencrypted BitLocker header (as explained in more detail in \ref{sec:case_study:tpm_only}).
After setup and during each boot, BitLocker relies on the \ac{tpm} to \emph{unseal} (\ref{subsubsec:sealing}) the \ac{vmk}. 
However, the \ac{tpm} will only do so if the \ac{pcr} register values match those saved in secure storage and defined during the BitLocker setup.
\\
This makes the \emph{TPM(-only)} protector entirely transparent for the user during boot, as it requires no additional user interaction when the \acp{pcr} are in the expected state.
If not, BitLocker will fall back on the recovery protector.
In order to decrypt it, it will prompt the user to enter the recovery key.

\subsection{AMD Secure Processor}
\label{sec:background:amd_sp}

The \acf{amd-sp} is a dedicated security co-processor part of all recent AMD \acp{soc}, including the Ryzen CPUs.
Since its introduction in 2013, the \ac{amd-sp} (formerly known as \ac{psp}) has acted as the \acf{rot} of the \ac{soc} \cite{lai_amd_2013}.
Its responsibilities on the Ryzen platform include the early \ac{soc} initialization, starting and initializing the secure boot chain, providing a \ac{tee}, and hosting the \ac{ftpm} application \cite{advanced_micro_devices_whitepaper_2021, cohen_full_2018, buhren_all_2020}.

\subsubsection{Early Boot}
\label{sec:background:amd_sp:boot}

\begin{figure*}[t]
  \includegraphics[width=\textwidth]{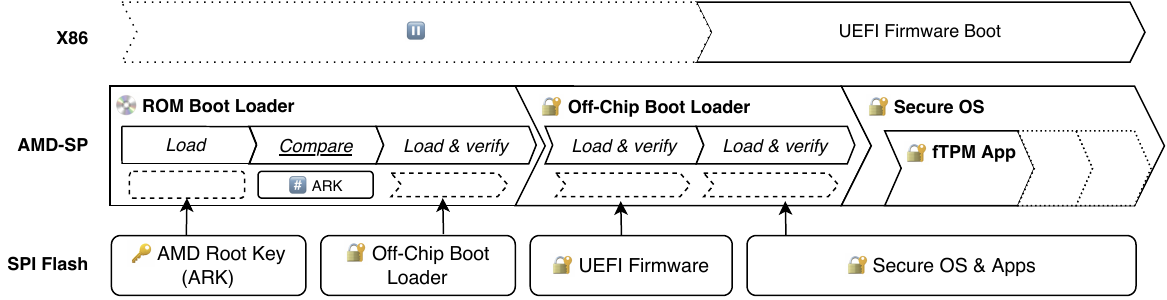}
  \caption{Early boot of a modern AMD CPU. A key denotes a public key, a pound sign a hash, and a lock a signed element.}
  \label{fig:amd_boot}
\end{figure*}

The \ac{amd-sp} boots before the main X86 cores of the AMD \ac{soc} \cite{buhren_uncover_2019}.
As illustrated in Figure \ref{fig:amd_boot}, the first boot stage to run on the \ac{amd-sp} is an immutable \ac{rom} boot loader.
After some minimal system initialization, the \ac{rom} boot loader loads the \acf{ark} from the \ac{bios} flash chip and verifies the key by comparing its SHA256 digest to a known value that is part of the \ac{amd-sp}'s immutable \ac{rom} \cite{buhren_one_2021}.
Once verified, the \ac{amd-sp} loads its next boot stage, the so-called \emph{off-chip} boot loader, from the \ac{bios} flash chip and verifies it using the \ac{ark}.
Finally, the \emph{off-chip} boot loader initializes various components of the \ac{soc} and executes other system initialization routines like DRAM training.

After the system initialization, the \ac{amd-sp} loads the initial boot firmware into X86 memory and starts the X86 cores \cite{buhren_all_2020}.
Once the X86 cores are running, the \emph{off-chip} boot loader is replaced with a microkernel. 
This microkernel acts as a \ac{tee} for, among other applications, the \ac{ftpm} implementation.

\subsubsection{Firmware}
\label{sec:background:amd_sp:firmware}

Besides the ROM boot loader, all firmware executed on the \ac{amd-sp} is loaded from \ac{bios} flash.
Although the general format is standardized by the \ac{uefi} volume standard, the \ac{amd-sp}'s code is stored in a proprietary file system.
The open-source tool \emph{PSPTool} \cite{werling_psptoolpsptool_2022} aims to parse and modify these \emph{firmware file systems}.
It has been used by Buhren et al.'s security-related publications about AMD processors \cite{buhren_insecure_2019, buhren_all_2020, buhren_one_2021}.

\subsubsection{Cryptographic Co-Processor (CCP)}
\label{sec:background:amd_sp:ccp}
The \ac{amd-sp} features a dedicated hardware component, the so-called \acf{ccp}, that allows offloading various cryptographic operations and is available to both the \ac{amd-sp} and the X86 CPU.
Although not officially documented by AMD, the CCP Linux driver gives an insight into its functionality \cite{advanced_micro_devices_ccp-devc_2017}.
The \ac{ccp} features a local memory space commonly called the \acf{lsb}, which can hold keys or other data for cryptographic operations.

\subsubsection{Attacks}
\label{sec:background:amd_sp:attacks}
\label{sec:background:amd_sp:fi_attack}

\label{subsubsec:rom_bl_attack}
In \cite{buhren_all_2020}, Buhren and Eichner present their emulation efforts of the \ac{psp} and disclose a critical vulnerability in the ROM boot loader of supposedly all Zen 1 and Zen Plus CPUs.
A stack-based buffer overflow caused by user-provided data from the \ac{spi} flash yields privileged code execution in the \emph{ROM Boot Loader} (as seen in Figure \ref{fig:amd_boot}).
Unfortunately, since the vulnerability lies in the on-chip \ac{rom} of the \ac{psp}, AMD cannot issue any fix for this vulnerability.

However, in this paper, we use a voltage fault injection attack to gain code execution on the \ac{amd-sp} of the newer Zen 2 and Zen 3 CPU generations as introduced by Buhren et al. in \cite{buhren_one_2021}.
This attack leverages the \ac{svi2} bus, allowing the AMD \ac{soc} to update its supply voltages dynamically.
By injecting packets onto this bus, an attacker causes a short drop in the \ac{amd-sp}'s supply voltage and induces a fault in the \ac{amd-sp}.
With a carefully timed fault injection, Buhren et al cause the \emph{Compare} operation illustrated in Figure \ref{fig:amd_boot} to accept a modified \ac{ark} previously placed on the \ac{bios} flash chip.
With \emph{PSPTool's} capabilities to replace and resign various \ac{amd-sp} firmware components, this fault injection attack can be used to gain code execution in various stages of the \ac{amd-sp}'s runtime.

\section{Related Work}
\label{sec:related_work}

\subsection{TPM Attacks}

\subsubsection{Side-channel attacks against fTPMs and dTPMs}
\label{subsubsec:side_channel_attacks}
The most recent academic attack on \acp{tpm} is \cite{moghimi_tpm-fail_2020}.
Moghimi et al. perform a black-box timing analysis of \ac{tpm} 2.0 devices and find secret-dependent execution times during signature generation.
These timing leakages are discovered on both an Intel \ac{ftpm} and a \ac{dtpm} by STMicroelectronics.
\\
Both Intel and STMicroelectronics have released firmware updates addressing the vulnerabilities \cite{zorz_intel_2019}.

\subsubsection{Power management attacks against the TPM 2.0 specification and tboot}
\label{subsubsec:power_management_attacks}
In \cite{han_bad_2018}, two sorts of \ac{tpm} attacks regarding power management are reported where Han et al. find a way to reset and forge the \ac{tpm}'s \acp{pcr} values.
One vulnerability targets a grey area design flaw in the \ac{tpm} 2.0 specification.
The other exploits an implementation flaw in the most popular measured launch environment used with Intel's \ac{txt}, `tboot`.
\\
While the authors provided a patch to mitigate the latter, they contacted and reported their findings of the former to Intel, Dell, Gigabyte, and Asus.

\subsubsection{AMD fTPM trustlet code execution attack}
In an earlier public disclosure of AMD's \ac{ftpm} security \cite{cohen_full_2018}, Cohen finds a stack-based buffer overflow in the \ac{ftpm} trustlet running on the \ac{psp}.
The vulnerability is exploitable through user-controlled data in a \ac{tpm} 2.0 call and allows full control of the program counter.
By applying additional exploit techniques like return-oriented programming, this vulnerability can make it possible to break the \ac{ftpm}'s security guarantees.
\\
AMD issued firmware updates to mitigate the vulnerability \cite{advanced_micro_devices_amd_2018}.

\subsubsection{LPC sniffing attacks against dTPMs}
\label{subsubsec:lpc_sniffing_attacks}
Even though \acp{dtpm} are tamper-resistant devices, the LPC bus connecting it to the main CPU is not.
In \cite{winter_hijackers_2013}, Winter and Dietrich show that passive attacks against \acp{dtpm}'s bus communication can be mounted with reasonably inexpensive equipment.
Moreover, active attacks allow the authors to circumvent any security mechanism provided by the \ac{tpm}, e.g., the chain of trust.

\subsection{BitLocker Attacks}
\label{sec:related_work:bitlocker_attacks}

\noindent Two of these attacks remain problematic for BitLocker until today:
\begin{enumerate}[itemsep=-.5em]
\item Microsoft recommends defeating any \emph{power management-based attacks} (\ref{subsubsec:power_management_attacks}) by \enquote{disabl[ing] Standby power management, and shut[ing] down or hibernat[ing] the device before it leaves the control of an authorized user} \cite{microsoft_bitlocker_2021-1}.
\\
\item To protect against any \emph{LPC sniffing attacks against dTPMs} (\ref{subsubsec:lpc_sniffing_attacks}), Microsoft advises using a TPM with PIN protector instead of a purely \ac{pcr}-based TPM protector \cite{microsoft_bitlocker_2021-1}. 
\end{enumerate}

\subsubsection{Cold boot attack against RAM}
\label{subsubsec:cold_boot_attacks}
\emph{Cold boot attacks} are a physical side-channel attack in which an attacker performs a memory dump of the RAM by performing a hard reset of the target machine.
They rely on the data remanence property of DRAM and SRAM that allows retrieving memory contents seconds to minutes after power off.
Since BitLocker stores the essential key material in memory, this attack can be mounted regardless of the used BitLocker protectors.
\\
Full memory encryption, as implemented by AMD \cite{advanced_micro_devices_amd_2020} and proposed by Intel \cite{intel_corporation_whitepaper_2017}, can potentially mitigate cold boot attacks.

\subsubsection{Drive-by DMA attacks}
In the past, BitLocker was a popular target for \ac{dma} attacks using FireWire \cite{bock_firewire-based_2009}.
As it allows an attacker to retrieve BitLocker keys directly from memory, its nature is similar to \ref{subsubsec:cold_boot_attacks} but with less complex hardware requirements.
Such attacks are still relevant \cite{ruytenberg_breaking_2020} but actively mitigated by leveraging the system IOMMU to implement kernel \ac{dma} protection \cite{microsoft_kernel_2021}. 

\subsubsection{Dictionary attacks}
Decrypting the BitLocker \ac{vmk} was optimized for GPUs in \cite{agostini_bitcracker_2022}.
However, cracking it in such a fashion remains a costly and time-consuming task.

\section{fTPM Attack}
\label{sec:attack}

In this section, we present our attack on AMD's proprietary fTPM, which allows us to decrypt sealed TPM objects regardless of their authorization policy (\ref{subsubsec:sealing}).

To carry out our attack, we execute a small payload that leaks a \emph{chip-unique secret} from the CPU (\ref{sec:attack:secret_extraction}).
This secret is used to derive the encryption and signature keys for the fTPM's \emph{non-volatile data}, which is stored on the BIOS flash chip (\ref{sec:attack:nv_storage}).
As a result, we now have the ability to decrypt or modify the fTPM's \emph{non-volatile} state, which we use to get access to the \emph{storage keys} of sealed TPM objects (\ref{sec:attack:tpm_object}).

\subsubsection*{Hardware Access}

In order to analyze and attack AMD's fTPMs, we use the \emph{voltage fault injection attack} presented by Buhren et al. \cite{buhren_one_2021}.
This attack (\ref{sec:background:amd_sp:fi_attack}) allows us to gain code execution during various stages of the \ac{amd-sp}'s firmware, including the \ac{ftpm} application (\ref{sec:background:amd_sp:boot}).
The attack requires access to the motherboard of the target system (\ref{sec:attack:fi_attack}), particularly its SPI bus and voltage regulators.
After leaking a CPU's \emph{chip-unique secret}, no more glitching is required.
To attack this CPU, we now merely need access to the BIOS flash, which can simply be read by accessing the motherboard's SPI bus.

\subsubsection*{Older CPUs}

Our attack targets AMD Ryzen CPUs of the microarchitecture generations Zen 2 \& 3, which use a common fTPM implementation.
The older Zen (1) and Zen $+$ CPUs use a different fTPM implementation and, particularly a different non-volatile storage format.
If an attacker reverse engineers this storage format and its key-derivation algorithm, the same attack approach does apply.
Since there are code-execution attacks for these CPUs that do not need fault-injection whatsoever (\ref{subsubsec:rom_bl_attack}), they should be considered even more vulnerable.

\subsection{Voltage Fault Injection Attack}
\label{sec:attack:fi_attack}

\begin{figure}
  \centering
  \includegraphics[width=\columnwidth]{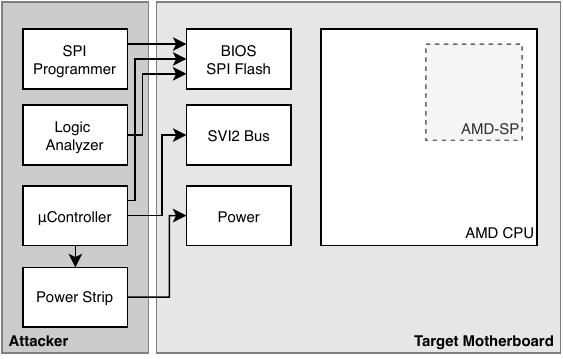}
	\caption{Physical connections necessary for the attack}
  \label{fig:setup_schematic}
\end{figure}

Recall the \ac{amd-sp}'s boot-process (\ref{sec:background:amd_sp}):
An immutable ROM bootloader loads and verifies the \acf{ark}, which is used to verify all further firmware components, including the \emph{off-chip bootloader} that is executed after the ROM bootloader.
Buhren et al.'s \emph{voltage fault injection attack} causes the \ac{amd-sp} to accept an invalid \ac{ark} and thus enables an attacker to replace and resign various firmware components on AMD Epyc CPUs.
In this section, we introduce details about this attack and highlight changes that were necessary to apply the attack to Ryzen CPUs and laptops.

\begin{table}[h]
  \centering
  \normalsize
  \begin{tabular}{lr}
  Description & Cost in USD \\ \hline
	Teensy 4.0 Development Board & $\sim 20 \$$ \\
	\ac{spi} flash programmer & $\sim 15 \$$ \\
	Logic Analyzer & $\sim 15 \$$ \\
	Driver IC and additional resistors & $\sim 5 \$$ \\
	Flash chip test clip & $\sim 5 \$$ \\
	Test probes with spring-loaded pins & $\sim 120 \$$ \\
	Controllable Power Relay & $\sim 15 \$$ \\ \hline
	Total & $\sim 195 \$$
  \end{tabular}
  \vspace{1em}
	\caption{Total hardware costs for the attack at time of writing}
  \label{tbl:costs}
  \vspace{-2em}
\end{table}

\subsubsection{Hardware Setup}
\label{sec:attack:fi_attack:hardware}

To leverage the \ac{amd-sp}'s susceptibility towards \emph{voltage fault injection}, an assortment of readily available hardware is required:

\begin{description}[leftmargin=.5cm]
  \item[SPI Programmer] In order to read and write the \ac{bios} flash chip, an \ac{spi} flash programmer is needed.
  \item[Attack µController] A small microcontroller board injects the \emph{voltage fault} and generates the fault injection trigger.
  As described by Buhren et al. \cite{buhren_one_2021}, a driver IC and some additional resistors aid the microcontroller with its \ac{svi2} bus injection.
  \item[Reset Method] Since the attack also requires frequent restarts of the device under attack, the microcontroller needs to be able to reset the target.
  In the case of a desktop motherboard, the original glitch attack's method using the ATX case header's reset pin can be used \cite{buhren_one_2021}.
  For laptops, which generally do not expose an ATX case header, a relay can be used to control the laptop's power supply while disconnecting any batteries.
  \item[Logic Analyzer] Buhren et al. use the \ac{spi} bus to extract data from the \ac{amd-sp} \cite{buhren_one_2021}.
  To capture this data, a logic analyzer is also connected to the \ac{spi} bus and records the communication.
\end{description}
Figure \ref{fig:setup_schematic} illustrates all necessary connections to the target motherboard.

\begin{figure}
  \includegraphics[width=\columnwidth]{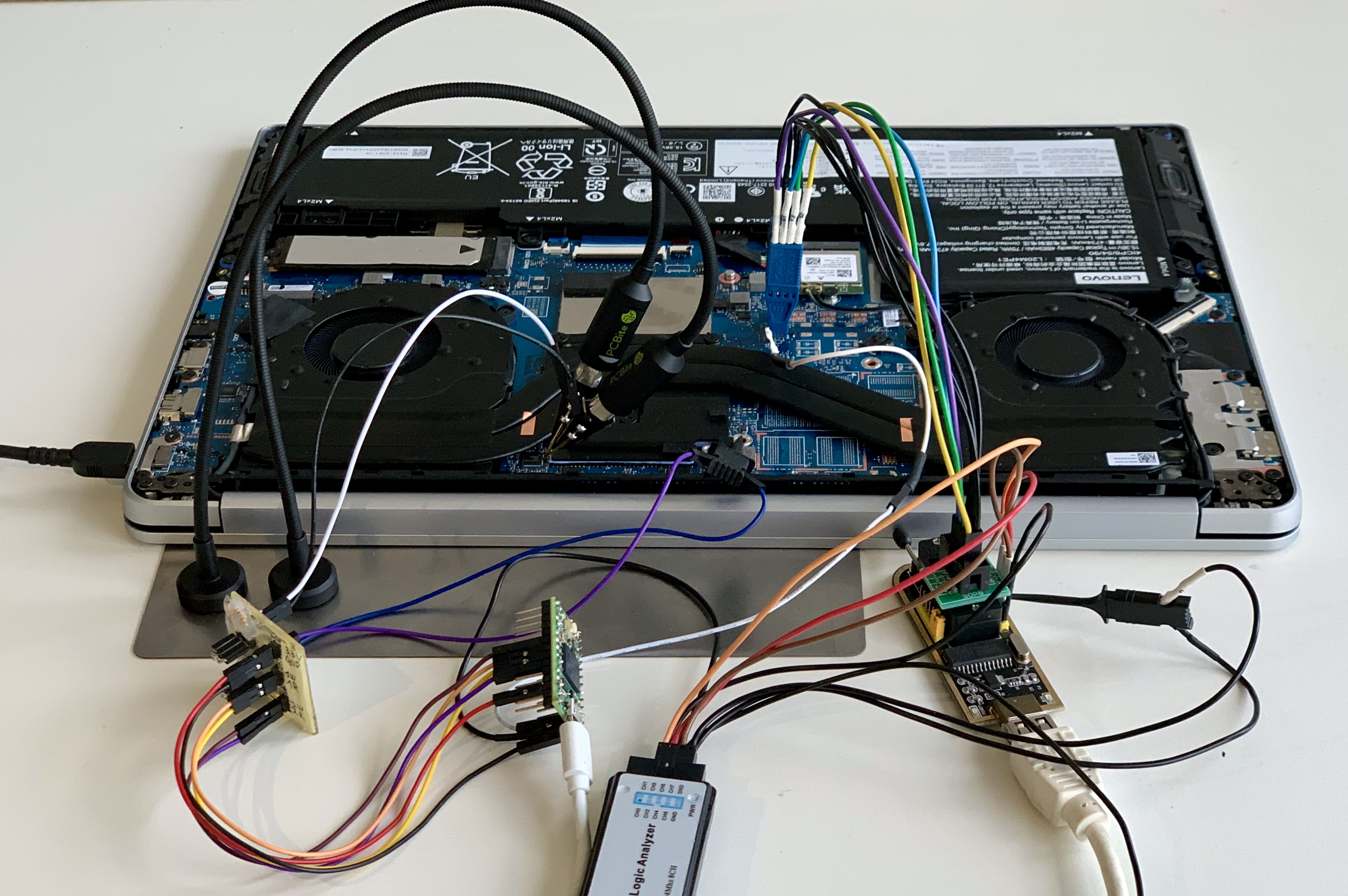}
	\caption{Attack setup on a Lenovo Ideapad 5 Pro-16ACH6}
  \label{fig:setup_photo}
\end{figure}

For the additional connections to the motherboard's \ac{svi2} bus, we used test probes with spring-loaded tips.
With these, we could reliably connect our injection hardware without soldering, even to adjacent pins of the TQFN packages used by all target \acp{vr} that we attacked.
\\
The final attack setup can be seen in Figure \ref{fig:setup_photo}.
At the time of writing, the total hardware cost amounts to under 200 USD, more than half of which is spent on the test probes (Table \ref{tbl:costs}).

\subsubsection{Gaining Code Execution}
\label{subsec:code_exec}

Once the fault injection hardware is connected, the attack by Buhren et al.\ consists of a manual \emph{parameter determination} phase and a brute-force search for a final \emph{delay} parameter \cite{buhren_one_2021}.
The first step requires around 30 minutes of manual attention at the moment, we believe it can be automated, as algorithm-like descriptions of the \emph{parameter determination} process can already be found in the attack's supplementary code repository \cite{buhren_one_2021, buhren_glitching_2021}.
The attack's second phase consists of a loop of repeated attack attempts to search for the last to-be-determined parameter and execute the attack's payload.

\subsubsection{Payload creation}
\label{sec:attack:fi_attack:payload}

Our payloads consist of a short \emph{ARMv7a} assembly section (the \ac{amd-sp}'s architecture \cite{lai_amd_2013}) to bootstrap our C-code payload.
The necessary hardware details, e.g., the \acs{mmio} interfaces of the \ac{amd-sp}'s \ac{spi} controller or the \ac{ccp} (\ref{sec:background:amd_sp:ccp}), can be found in Buhren et al.'s supplementary repository \cite{buhren_glitching_2021}.
In order to finally gain code execution, we create our own RSA key pair and replace the \ac{ark} on the \ac{bios} flash chip with our key.
We then replace the off-chip boot loader on the \ac{bios} flash chip and resign it with our \ac{ark} replacement key.
Once we successfully inject a fault and the \ac{amd-sp} accepts our \ac{ark}, it will load and execute the modified off-chip boot loader.

To make cryptographically consistent changes to the \ac{amd-sp}'s file system, we rewrote parts of the open-source tool \emph{PSPTool} \cite{werling_psptoolpsptool_2022} that check and re-create signatures when a key or other firmware file is replaced.
Our modifications to \emph{PSPTool} have since been upstreamed to the current version of the tool \cite{werling_psptoolpsptool_2022}.
In addition to executing a payload instead of the \emph{off-chip boot loader}, we can also patch existing firmware components, resign the image, and boot the whole AMD \ac{soc} with the modified firmware.

\subsection{Non-Volatile Storage}
\label{sec:attack:nv_storage}

The \acf{nv} state of AMD's fTPM is stored on the motherboard's \ac{bios} flash chip.
To protect against an attacker with read or write capabilities to the \ac{bios} flash chip, the confidentiality and integrity of the NV state are cryptographically protected.
We reverse-engineered the data structures of these files, as well as the storage and integrity key derivation algorithms for Ryzen CPUs of the Zen 2 and Zen 3 microarchitecture generations\footnote{This storage format differs for Zen 1/+-based systems.}.

\begin{figure}
  \includegraphics[width=\columnwidth]{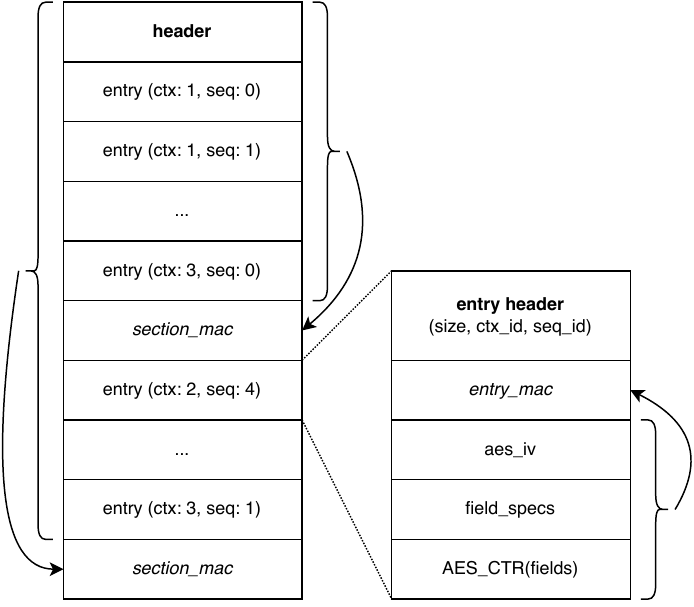}
  \caption{Format of an NV section (left) and entry (right)}
  \label{fig:nv_storage}
\end{figure}

The fTPM's non-volatile state can be found in a file stored alongside the \ac{amd-sp}'s firmware on the \ac{bios} flash chip.
\emph{PSPTool} labels this file \texttt{NV\_DATA}.
Although the \texttt{NV\_DATA} file's sensitive data is encrypted, its metadata and structure can be understood without access to the encryption keys.
\texttt{NV\_DATA} files are divided into two \SI{64}{\kibi\byte} sections featuring append-only data structures (left side of Figure \ref{fig:nv_storage}).
Once both sections are filled, the older of the two sections is overwritten and it is ensured that the new file contains all the data necessary for the \ac{ftpm}.
Each section consists of a header and multiple entries, each of which is associated with a \emph{context} that indicates the entry's usage.
Furthermore, the entries of each context are ordered by an increasing \emph{sequence} number.

Each entry consists of a variably sized body encrypted using AES128 in counter mode.
The entry's body can be subdivided into up to seven variably sized fields.
The integrity of each entry is protected by an HMAC-SHA256 \acs{mac} over the encrypted body, the \acs{iv} for the encryption cipher, and the unencrypted field-length specification (see right side of Figure \ref{fig:nv_storage}). 
In addition to this \emph{Encrypt-then-MAC} protection of each entry, the section data is protected by occasional HMAC-SHA256 \acp{mac} over the whole section contents up to the latest entry (left side of Figure \ref{fig:nv_storage}).
A tool for parsing (and decrypting) this \ac{nv} storage format is presented in Section \ref{sec:attack:nv_storage:tooling}.

\subsubsection{NV storage key derivations}
\label{sec:attack:nv_storage:key_derivation}

\begin{figure}
  \centering
  \includegraphics[width=0.8\columnwidth]{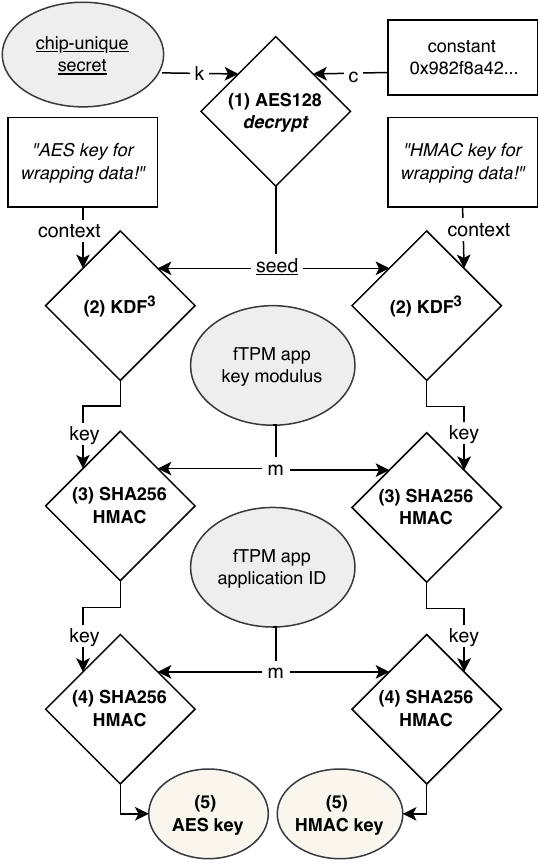}
  \caption{Key derivation used for the NV data keys}
  \label{fig:key_derivation}
  \vspace{-1.5em}
\end{figure}

To reverse-engineer the key deri\-vation process for both the \emph{storage (AES) key} and the \emph{integrity (HMAC) key}, we used the ability to patch and replace arbitrary firmware components of the \ac{amd-sp} (\ref{sec:attack:fi_attack}).
The \ac{ftpm} application runs as an application in the \emph{SecureOS} microkernel (\ref{sec:background:amd_sp:boot}).
Another application, labeled \texttt{DRIVER\_ENTRIES} by the \emph{PSPTool}, implements drivers for device-specific functionality, like the \ac{ccp} or \ac{spi} bus.
We statically analyzed the \texttt{DRIVER\_ENTRIES} binary and created a modified version that logs every cryptographic operation, including its inputs and outputs (see Figure \ref{fig:key_derivation}), to the \ac{amd-sp}'s \ac{spi} bus.

The storage and integrity keys are derived from a \SI{128}{\bit} secret unique to each CPU.
This chip-unique secret is held by the \ac{ccp} of the \ac{amd-sp}.
It is present at address zero of the \ac{lsb} (\ref{sec:background:amd_sp:ccp}).
Across all devices we tested, the derivation process (illustrated in Figure \ref{fig:key_derivation}) was consistent:
\begin{enumerate}
    \item The secret is used as key in an \emph{AES128} decrypt operation with a constant as ciphertext.
    \label{nv_key_derivation:1}
    \item From the AES operation's output, two \SI{256}{\bit} values are derived using a NIST specified \ac{kdf}\footnote{
      \ac{kdf} in counter mode with HMAC-SHA256 as \emph{pseudorandom function} and an empty context, specified in NIST's SP 800-108 \cite{chen_recommendation_2009}.
    }, where ``AES key for wrapping data'' and ``HMAC key for wrapping data'' are used as the \emph{label} inputs.
    \item Into each of these values, the signing key of the \ac{ftpm} application is mixed.
    This is done by computing the SHA256 digest of the RSA key's modulus and calculating the HMAC-SHA256 of this digest with the secret as a key.
    \item Similarly, the \emph{id} under which the \ac{ftpm} application is run in the \emph{SecureOS} is mixed in.
    The resulting values are the HMAC-SHA256 of the \SI{128}{\bit} id, used as message, and the respective secrets, as keys.
    \item Finally, the first value is truncated to its first \SI{128}{\bit}s and used as the \emph{AES storage key}, while the full \SI{256}{\bit}s of the second value are used as the \emph{HMAC integrity key}.
\end{enumerate}

\subsection{Secret Extraction}
\label{sec:attack:secret_extraction}
\label{subsubsec:extracting_cpu_secrets}

We extracted the output of each step of the key derivation process by booting the system with a patched \texttt{DRIVER\_ENTRIES} binary and analyzing the traced cryptographic operations.
However, this entails the challenge of leaving the boot functionality intact.

In order to remove the need to fully boot the system and to make our attack work with other versions of the \texttt{DRIVER\_ENTRIES} binary, we built a payload (\ref{sec:attack:fi_attack:payload}) that directly computes the output of step (\ref{nv_key_derivation:1}) of the key derivation process and writes the result to the \ac{spi} bus.
Then, using the fault injection attack, we can execute the payload in place of the off-chip boot loader (\ref{sec:attack:fi_attack}) and, with the logic analyzer, extract the seed value (underlined in Figure \ref{fig:key_derivation}) from the \ac{spi} bus.

This extracted seed is all that is necessary to decrypt a Zen 2 or Zen 3-based \ac{ftpm}'s internal state from a \ac{bios} image.
However, we additionally provide the means to leak the unmodified \emph{chip-unique secret} from pre-Zen 3 CPUs:

\subsubsection{Extracting the chip-unique secret}

The chip-unique secret (underlined in Figure \ref{fig:key_derivation}) used in the \ac{ftpm}'s key derivation is contained in a read-protected area of the \ac{ccp} (\ref{sec:background:amd_sp:ccp}).
This is presumably meant to add a layer of protection by preventing the secret from being extracted from the \ac{lsb}.
But, due to an oversight in the \ac{ccp}'s interface design, we are able to extract the entire contents of the \ac{lsb} for Zen 1, Zen +, and Zen 2 CPUs, including read-protected areas.
This extraction is possible due to the allowed use of unaligned key addresses in AES crypto operations.

\begin{figure}
  \centering
  \includegraphics[width=0.85\columnwidth]{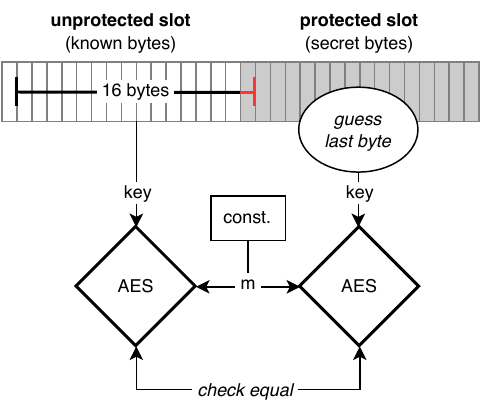}
  \caption{Our \ac{lsb} extraction technique illustrated}
  \label{fig:lsb_extraction}
  \vspace{-1.5em}
\end{figure}

Usually, the \ac{lsb} is accessed in multiples of 16 byte regions, so-called \emph{slots}, but the \ac{ccp} allows AES operations with arbitrarily aligned keys.
In particular, the \ac{ccp} allows us to execute an AES128 operation with a key that is contained partially in a read-protected and partially in an unprotected slot.
We use this to execute an AES128 encrypt operation with a key of which \SI{120}{\bit}s (15 bytes) are known, and only \SI{8}{\bit}s (1 byte) are unknown.
Furthermore, we can choose the operation's input and have access to its output.
This enables us to brute-force the unknown byte by comparing the AES operation's output to the 256 possible outputs that an AES operation with the given input and possible key values can have (see Figure \ref{fig:lsb_extraction}).
By shifting the key-window one byte at a time and repeating the single-byte brute-force attack, we can extract the entire read-protected slot.

On Zen 3 CPUs, AES operations with an unaligned key result in an error, foiling this extraction method for AMD's latest generation Ryzen CPUs.
Further inquiries into the CCP interface did not yield similar vulnerabilities for Zen 3.
Nevertheless, since the ciphertext for operation (1) of the key-derivation algorithm (Figure \ref{fig:key_derivation}) is constant, leaking the output of this operation, as described in Section \ref{sec:attack:secret_extraction}, is sufficient to achieve the same goal on Zen 3 CPUs.

\subsubsection{Tooling}
\label{sec:attack:nv_storage:tooling}

We included the payloads that implement both methods of extracting key derivation secrets in the paper's supplementary code repository \cite{jacob_pspreverseftpm_attack_2023}.
Additionally, to parse and decrypt the \ac{ftpm}'s \ac{nv} storage, we developed the Python tool \emph{amd-nv-tool}, whose code we also published alongside the paper.
\emph{amd-nv-tool} parses the unencrypted structure and metadata of the \texttt{NV\_DATA} file, derives the storage and integrity keys, and finally outputs the \ac{nv} storage's contents in a JSON representation.

\subsection{TPM Object Decryption}
\label{sec:attack:tpm_object}

At this point, we have an attack primitive to extract a chip-unique secret for a CPU and decrypt the \ac{ftpm}'s non-volatile storage with this secret.
One application of this decrypted \ac{tpm} state is to decrypt sealed \ac{tpm} objects (\ref{subsubsec:sealing}).
Externally stored (sealed) \ac{tpm} objects consist of a public and a private part.
The public part communicates metadata about the object, e.g., the object authorization policy, and uniquely identifies the object (\texttt{TPM2B\_PUBLIC} in \enquote{Part 2: Structures} of the TPM specification \cite{trusted_computing_group_trusted_2019}).
No part of this public part is encrypted or its integrity protected.

\begin{figure}
  \centering
  \hspace{2em}\includegraphics[width=0.9\columnwidth]{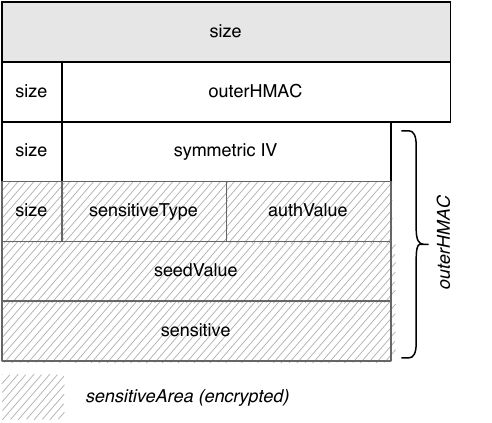}
  \caption{Format of \texttt{TPM2B\_PRIVATE} object}
  \label{fig:tpm_object}
  \vspace{-1.3em}
\end{figure}

In contrast, the object's private portion is protected with an \emph{encrypt-then-MAC} approach involving a \emph{symmetric} and an \emph{HMAC} key (see Figure \ref{fig:tpm_object}).
The encrypted part of the object consists of structural metadata, an authentication value, a seed, and the object's sensitive data.
The authentication value is used for PIN or passphrase authentication, while the seed adds entropy to the object and can be used to seal child objects of this object.
In the sensitive data part, we find the sealed data or, in the case that our object is a key, its symmetric or private key.

\subsubsection{Deriving the sealing keys}
\label{sec:attack:tpm_object:deriving}

As described in Section \enquote{23 Protected Storage Hierarchy} of \enquote{Part 1: Architecture} of the TPM specification \cite{trusted_computing_group_trusted_2019}, \ac{tpm} objects are organized in a tree of objects, where each object is `sealed' by its parent.
This is realized by deriving the sealing keys of an object from its public portion together with the parent's seed value (see \enquote{22 Protected Storage} of \enquote{Part 1: Architecture} of the TPM specification \cite{trusted_computing_group_trusted_2019}).
The root, or \emph{primary}, objects of these trees can be deterministically generated from a persistent seed value but are, per default, cached in the \ac{tpm}'s non-volatile storage \cite{arthur_practical_2015}.

We used this caching to our advantage and implemented an unsealing tool that searches all consecutive \SI{256}{\bit} values in the decrypted non-volatile fTPM state for the correct \emph{primary} seed.
To check whether a possible seed value is the desired primary seed, our tool derives the corresponding HMAC key and tries to verify the HMAC of the object's private portion.
Once this verification succeeds, we use this \emph{primary} seed to derive the symmetric key and decrypt the object's private part.

This means, with access to the decrypted non-volatile \ac{ftpm} state, we are able to decrypt any \ac{tpm} object whose primary key is cached on the \ac{ftpm}, \emph{regardless} of the object's authentication policies.
We implemented the key derivation and decryption detailed above in our \texttt{amd\_ftpm2\_unseal} tool, published alongside this paper.

\subsection{Results}
\label{sec:attack:results}

To summarize, we consider an attacker in possession of a victim device, e.g., a laptop, wants to carry out the attack described in this paper to attack any \ac{ftpm}-based security measure.
The necessary steps they will have to mount are as follows:

\begin{enumerate}
\item Backup the BIOS flash image using an SPI flash programmer
\item Connect the fault injection hardware and determine the attack parameters (\ref{sec:attack:fi_attack})
\item Compile \& deploy the payload extracting the key derivation secret (\ref{sec:attack:secret_extraction}) 
\item Start the logic analyzer to capture the extracted key derivation secrets via SPI
\item Start the attack cycle on the target machine until the payload was executed successfully 
\item Parse \& decrypt the NVRAM using the BIOS ROM backup and payload output with \emph{amd-nv-tool}
\item Extract and decrypt TPM objects protected by this \ac{ftpm} with \emph{amd\_ftpm\_unseal}
\end{enumerate}

Note that once the seed or chip-unique secret has been extracted (steps 2 to 5), the attack can be re-mounted quickly by reading the BIOS flash image, parsing the contained NVRAM, and decrypting external TPM objects (steps 1, 6, and 7).
This process takes us about 15 minutes on a standard laptop.

\begin{table}
  \centering
  \small
  \begin{tabular}{ l l l }
     \textbf{System}             & \textbf{CPU}         \\ \hline
     Lenovo Ideapad 5 Pro 16ACH6 & Ryzen 5600H (Zen 3)         \\
     Asrock 520M-HDV             & Ryzen 5600X (Zen 3)         \\
     Asrock 520M-HDV             & Ryzen 3600 (Zen 2)
  \end{tabular}
  \vspace{.6em}
  \caption{Targets verified to be vulnerable to our \ac{ftpm} attack}
  \label{tbl:targets}
  \vspace{-1.5em}
\end{table}

We successfully executed the attack end-to-end on the hardware listed in Table \ref{tbl:targets}.
After having gained experience with the attack, we are able to perform the full attack on a new device within two to three hours.
All necessary tools, as well as sample intermediate data, are available in the supplementary repository \cite{jacob_pspreverseftpm_attack_2023}.

Furthermore, we were able to mount the \ac{lsb} secret extraction (\ref{sec:attack:secret_extraction}) on Zen 1 and Zen + CPUs by using the less complex ROM boot loader attack (\ref{subsubsec:rom_bl_attack}).
A simplified attack (without the need for voltage glitching) can, therefore, compromise Zen 1/+-based \acp{ftpm}, if an attacker conducts the necessary reverse engineering efforts.

\section{Case Study: BitLocker}
\label{sec:case_study}

To evaluate the severity of our \ac{ftpm} attack on \ac{tpm}-based applications, we conducted a case study on the implementation of \emph{BitLocker's} \emph{TPM-only} and \emph{TPM and PIN} protectors (\ref{sec:background:bitlocker}).
Since a \emph{TPM-only} strategy does not apply any other means of protection than the \ac{tpm}, we expect all such protection mechanisms to be broken by our attack.

However, as shown in Figure \ref{fig:available_protectors}, BitLocker provides additional authentication factors which can be combined with a TPM: \emph{PIN}, i.e., a 4- to 20-digit numeric or alphanumeric passphrase, and \emph{StartupKey}, storing part of the encryption key on a USB flash drive.
Since the \emph{StartupKey} protector, according to Microsoft documentation, does not make any use of the \ac{tpm} \cite{microsoft_prepare_2021} but supposedly acts as another layer of cryptographic protection, it does not apply to our \ac{ftpm} attacks and is not subject to our case study.

\subsection{Threat Model}
\label{sec:case_study:threat_model}

BitLocker \enquote{addresses the threats of data theft or exposure from lost, stolen, or inappropriately decommissioned computers} \cite{microsoft_bitlocker_2021-2}.
Since all these events regard the \emph{physical} state of the protectee's device, BitLocker's goal is to secure data from an attacker with \emph{physical} access to the device.
However, Microsoft differentiates two types of (physical) attackers:
\begin{enumerate}
\item An \textbf{opportunistic attacker} \enquote{does not use destructive methods or sophisticated forensics hardware/software} [...] \enquote{Physical access may be limited by a form factor that does not expose buses and memory.}
\item A \textbf{targeted attacker} has \enquote{plenty of time; this attacker will open the case, will solder, and will use sophisticated hardware or software.} 
\end{enumerate}

Our threat model considers a targeted attacker who has gained prolonged access to the device, e.g., a thief targeting a company laptop.
This attack type is commonly referred to as an \emph{Evil Maid attack}.
However, as we will discuss in detail in Section \ref{sec:discussion:requirements:zen1}, for Zen 1 and Zen + systems, our attack may be suitable for an opportunistic attacker as well.

\subsection{TPM-only Protector}
\label{sec:case_study:tpm_only}

To apply our attack described in Section \ref{sec:attack}, we first need to get ahold of the \ac{tpm} objects that facilitate \emph{BitLocker's} TPM protection.
The \emph{Dislocker} toolchain \cite{aorimn_dislocker_2022} can parse the unencrypted headers and metadata of a \emph{BitLocker} volume and decrypt a volume given one of the volumes \acp{vmk}.
For a \emph{BitLocker} volume protected by a \emph{TPM-only} protector, the metadata corresponding to that protector consists of a single \emph{datum} labeled \enquote{\texttt{TPM-encoded}}.
This \emph{datum} contains the sealed \ac{tpm} object containing a \ac{vmk} and the metadata required to fulfill the \ac{pcr} policy authentication of the object.

In detail, the datum contains the concatenation of the TPM object's private and public portion (\ref{sec:attack:tpm_object}), to which another data structure specifying the \ac{pcr} policy is appended.
As described in Section \ref{sec:attack:tpm_object:deriving}, with the decrypted non-volatile \ac{ftpm} state, our \texttt{amd\_ftpm2\_unseal} tool can extract the wrapping keys for the sealed object and decrypt its contents.
The unsealed \ac{tpm} object contains another \emph{BitLocker} \emph{datum}, the last \SI{256}{\bit}s of which are the disk's \ac{vmk}.
To verify our exploit, we mounted and decrypted the \emph{BitLocker} volume using the \texttt{dislocker} toolchain's \texttt{dislocker-fuse} command \cite{aorimn_dislocker_2022}.

\subsection{TPM and PIN Protector}
\label{sec:case_study:tpm_and_pin}

The \emph{TPM and PIN} protector consists of multiple \emph{datums} in a volume's BitLocker metadata.
These include \emph{datums} used for key stretching -- similar to those used for a recovery password -- in addition to a \ac{tpm} encoded \emph{datum} like the one of the \emph{TPM-only} protector.
In contrast to the \emph{TPM-only} protector, the unsealed \emph{datum} does not directly contain the \ac{vmk}, but an AES-CCM encrypted \emph{datum}.
\\
Once the \ac{tpm} protections have been circumvented, it is still necessary to brute-force, or otherwise acquire, the PIN.
We assume that BitLocker uses this scheme to provide reasonable protection even in case the \ac{tpm}'s security properties can be subverted:

\begin{displayquote}
  For some systems, bypassing TPM-only may require opening the case, and may require soldering, but it could be done at a reasonable cost.
  Bypassing a TPM with a PIN protector would cost much more, and require brute forcing the PIN.
  With a sophisticated enhanced PIN, it could be nearly impossible.
  \cite{microsoft_bitlocker_2021-1}
\end{displayquote}

Although recommended by Microsoft for some scenarios, we will discuss in \ref{sec:discussion:tpm_considerations:pin_setup} that it is not straightforward to step up from the default \emph{TPM-only} to a \emph{TPM and PIN} protector.

\subsubsection{Brute-forcing the PIN}
\label{sec:case_study:tpm_and_pin:brute_force}

The PIN-based key derivation involves a \SI{128}{\bit} salt and $1\,048\,576$ rounds of SHA256, which limit the speed of a brute-force attack to around $1\,000$ PINs/passwords per second on a GPU \cite{agostini_bitcracker_2022}.
In contrast, on a \ac{dtpm}, the hardware-aided anti-hammering mechanism limits the authorization rate to one attempt every ten minutes \cite{microsoft_trusted_2021}.

\begin{table}[h]
  \centering
  \normalsize
  \begin{tabular}{cclr}
  & & \multicolumn{2}{c}{time to brute-force} \\
  PIN/password & min-entropy & \ \ {\small fTPM} & {\small dTPM} \ \ \\
\hline
  4 digits & $2^9$ & $0.5$ sec & $3.5$ days \\
  10 digits & $2^{15}$ & $33$ sec & 7.3 mo \\
  10 characters & $2^{21}$ & $34$ min & $41$ yr \\
  20 characters & $2^{36}$ & $2.1$ yr & $1.3 \cdot 10^{6}$ yr
  \end{tabular}
  \vspace{1em}
  \caption{Estimated brute-force times based on NIST's password guessing entropy \cite{burr_electronic_2013}}
  \label{tbl:brute-force}
  \vspace{-1.5em}
\end{table}

As can be seen in Table \ref{tbl:brute-force}, the additional security provided by a numeric PIN -- in case of a compromised fTPM -- is negligible, while a 4-digit PIN already defeats a traditional evil maid attack if a dTPM is used.
On the other hand, a properly chosen passphrase can provide an adequate level of security even with a compromised fTPM.

\subsubsection{Alternative implementations}
\label{sec:case_study:tpm_and_pin_naive}

The naive approach to a \emph{TPM and PIN} protector is to only rely on the \ac{tpm}'s authentication mechanism to verify the \ac{pin}.
As detailed in Section \ref{sec:attack:tpm_object}, these authentication mechanisms do not mean the PIN/passphrase are involved in the sealed \ac{tpm}-object's encryption and therefore do not impose any restrictions on our attack.

An \ac{fde} tool that does not implement a \emph{TPM and PIN} strategy with a \emph{defense-in-depth} approach is the open-source tool \emph{systemd-cryptenroll}.
The \emph{systemd-cryptenroll} tool is part of the widely adopted \emph{systemd} project and acts as a management tool for encrypted disks conforming to the popular \emph{LUKS} standard \cite{systemd_systemd-cryptenroll_2022, systemd_systemd-cryptenroll_2022-1}.
Support for \ac{tpm} based protections has only been introduced recently and includes a \emph{TPM-only} and a \emph{TPM and PIN} strategy \cite{goronzy_cryptenroll_2022, poettering_cryptenroll_2020}.
Our analysis of the \emph{systemd-cryptenroll} code shows that a randomly generated \SI{256}{\bit} secret is directly sealed by the \ac{tpm}, protected either by a \ac{pcr} policy only or additionally a PIN.
The so-called \emph{LUKS} \emph{keyslot} (analogous to \emph{BitLocker's} \ac{vmk}) is then encrypted with the base64-encoded secret as passphrase.

Once the NV state is decrypted, the LUKS key is directly accessible and no brute-forcing is necessary.

To mitigate this issue, we recommend including the PIN in the passphrase protecting the \emph{LUKS} \emph{keyslot}.
With this approach, we can protect the disk and PIN with the brute-force resistant key-derivation mechanism of the \emph{LUKS} \emph{keyslot}, even if the \ac{tpm} encoded secret was leaked by an attack like the one described in this paper.
We proposed this approach to the maintainers of \emph{systemd-cryptenroll}.

\subsection{Results}
\label{sec:case_study:results}

Our case study shows that \emph{TPM-only} protection mechanisms for FDE are ineffective when the TPM's internal state can be extracted.
In particular, we demonstrated this with \emph{BitLocker's} \emph{TPM-only} protector using our attack against AMD's fTPM.
Furthermore, \emph{BitLocker's} \emph{TPM and PIN} protector retains the protection that a \emph{PIN-only} strategy would offer, which is basically negligible in the case of a numeric PIN.

With a passphrase, however, the same level of security that the \emph{passphrase-only} protector offers is retained, thanks to the additional layer of encryption applied to the \ac{vmk} before it is sealed by the \ac{tpm}.
This reveals a need for careful consideration when implementing a \emph{TPM and PIN} strategy.
On a vulnerable \ac{ftpm}, such a strategy without additional brute-force protection may be less secure against a targeted attacker than a \emph{PIN/Passphrase-only} strategy with a similarly strong PIN or passphrase.

\section{Discussion}
\label{sec:discussion}

In this section, we evaluate the feasibility and impact of our attack, propose potential mitigations, and, in light of publicly known TPM attacks, discuss important considerations for the use of BitLocker with a TPM, as well as the secure implementation of systems and applications relying on a TPM in general.

Our attack model assumes that an attacker has prolonged physical access to the target system.
Even though these are strong prerequisites, they are well within the threat model of typical applications using the \ac{tpm}, e.g., Microsoft BitLocker, detailed in Section \ref{sec:case_study:threat_model}.
For example, a typical scenario is when an attacker has stolen a laptop with valuable company secrets that is protected with \acf{fde}.

\subsection{Requirements}

In order to mount the key extraction, the attacker needs to be able to execute a custom payload on the \ac{amd-sp}.
Although we demonstrated the attack on newer Zen 2/3 systems, code execution on Zen (1)/+ systems can be achieved with a simpler setup.

\subsubsection{Zen 2/3 (Glitch attack)}

As detailed in Section \ref{sec:attack:fi_attack:hardware}, the hardware required to carry out the attack is readily available and amounts to under 200 USD.
\\
While the attack requires manual examination of each target motherboard, it has been adaptable to each target we tested in a few hours of work.
In addition, the use of spring-loaded pins has further removed the need for any soldering.
These capabilities correspond to a \emph{targeted attacker} defined by BitLocker's threat model (\ref{sec:case_study:threat_model}).
\\
Since the extracted secrets are chip-unique and immutable, an attacker can mount the glitching task, e.g., before the laptop reaches the end-user.
Afterward, they only need (software- or hardware-based) SPI flash reading capabilities to mount the remaining attack.

\subsubsection{Zen 1/+ (ROM attack)}
\label{subsubsec:requirements_zen1}

Due to an unpatchable vulnerability in the ROM boot loader of Zen 1 and Zen + CPUs \cite{buhren_all_2020}, a much simpler setup, with no voltage glitching required, can be used to extract the keys (\ref{subsubsec:rom_bl_attack}).
Instead, the attack only requires write capabilities for the \ac{bios} flash chip.
These can be achieved through either physical access with an \ac{spi} flash programmer or, in some cases, even through privileged \emph{software-only} access.
Although the proprietary NV storage format for Zen 1 and Zen + differs from the one we reverse-engineered, backporting `amd-nv-tool` would merely be diligence work.
The final attack would only involve opening the case and attaching the SPI flash programmer.
Thus, we argue that this would lift the requirements of this attack to be suitable even for an \emph{opportunistic attacker}.

\subsection{Capabilities}

There are two common \ac{fde} strategies employing a \ac{tpm} an attacker could face.
Depending on these strategies, our attack yields different capabilities for the attacker. 
Additionally, we discuss the capabilities an attacker gains against other TPM applications using our attack.

Our attack demonstrates that \ac{fde} using a \emph{TPM-only} protector, i.e., a protector sealed with a \ac{pcr} policy, can be decrypted regardless of the FDE implementation, which we verified by attacking Microsoft BitLocker (\ref{sec:case_study:results}).
\\
The attack's capabilities against an \ac{fde} secured by a \emph{TPM and PIN} protector rely heavily on the \ac{fde} implementation.
In particular, the anti-hammering protections of the \ac{ftpm} can be circumvented by our attack.
However, as shown in our case study (\ref{sec:case_study:tpm_and_pin}), a \emph{TPM and PIN} protector can be implemented such that the security of a \emph{PIN-only} strategy remains once the \ac{ftpm} state is compromised.
\\
This highlights the importance of BitLocker's \emph{TPM and PIN} protector and emphasizes the importance of the used PIN strength, which we will elaborate in Section \ref{subsubsec:bitlocker_effectively}. 
We discuss general considerations regarding \ac{fde} implementations in Section \ref{subsubsec:implement_fde_securely}.

In general, our attack gives an attacker access to the complete internal state of the \ac{ftpm}.
Since any authorization policy that specifies how and when a \ac{tpm} object can be used or accessed is ineffective (\ref{sec:attack:tpm_object}), it allows circumventing any security mechanism relying on the \ac{ftpm}. 
For example, an SSH private key protected by a \ac{tpm} (\ref{sec:background:tpm:applications}) would be vulnerable to our \ac{ftpm} state compromise.

\subsection{Mitigations}

Stopping arbitrary code execution attacks on the \ac{amd-sp} is the only way to mitigate our attack effectively. 
Unfortunately, this is currently not possible, as outlined in the following two Sections.
Nevertheless, starting with Section \ref{subsubsec:miti_lsb_extraction}, we propose soft measures that can make it harder to mount our attack.

\subsubsection{Zen 2/3 (Glitch attack)}

Since our attack is based on Buhren et al.'s \emph{voltage fault injection} attack \cite{buhren_one_2021}, it is not easily mitigable for currently available CPUs.
Mitigations involve changes in the hardware of the \ac{soc} or the ROM boot loader of the \ac{amd-sp} \cite{buhren_one_2021}.
They can therefore be expected at the earliest in the next microarchitecture generation.
This is an essential difference to attacks exploiting software vulnerabilities in the \ac{ftpm}'s code \cite{cohen_full_2018} or its execution environment \cite{moghimi_tpm-fail_2020}.
Such attacks can be (and have been \cite{advanced_micro_devices_amd_2018, zorz_intel_2019}) mitigated with updates to the relevant software components.

A notable development in the realm of hardware-based FI countermeasures is Intel's introduction of digital detection circuitry into their \ac{csme} \cite{nemiroff_tunable_2022, nemiroff_whitepaper_2022}, which is Intel's counterpart to the \ac{amd-sp}.
The \ac{csme} acts as the \ac{tee} for Intel's \ac{ftpm} implementation, called \ac{ptt} \cite{intel_corporation_code_2020}.
With this approach, Intel preemptively follows Buhren et al.'s mitigation recommendations.

\subsubsection{Zen 1/+ (ROM attack)}
\label{sec:discussion:requirements:zen1}

The ROM boot loader vulnerability presented in \cite{buhren_all_2020} cannot be mitigated in existing Zen 1/+ CPUs, as it lies in the read-only memory, but has been mitigated since.

\subsubsection{LSB secret extraction}
\label{subsubsec:miti_lsb_extraction}

On Zen 2 and below, we extracted the chip-unique secrets of the \ac{ccp} (\ref{sec:background:amd_sp:ccp}) through unaligned AES operations (detailed in Section \ref{subsubsec:extracting_cpu_secrets}).
These enabled us to derive the NV storage keys \emph{offline}.
While this extraction method cannot be mitigated on pre-Zen 3 CPUs, it has been mitigated in Zen 3 CPUs by disallowing the use of unaligned keys in AES operations.
However, since we can still perform the first stage of the NV storage key derivation \emph{online}, the measure could not mitigate our attack whatsoever.

\subsubsection{Limiting hardware access}

Our attack requires access to the \ac{spi} and \ac{svi2} buses of the target system.
Therefore, hardware obfuscation techniques could impede the attack, e.g., by hiding the relevant buses on the mainboard or making them otherwise physically inaccessible.
However, we believe that these techniques will not effectively mitigate but only delay attacks.
For example, if the \ac{svi2} was inaccessible, an attacker could directly interfere with the mainboard's passive power supply components.

\subsubsection{Software obfuscation}

In the same vein, AMD could change the NV storage layout or its key derivation algorithm, e.g., by changing its constant value (Figure \ref{fig:key_derivation}).
Such obfuscation attempts can be circumvented by additional reverse engineering.
In addition, on Zen 3, this would potentially also require extracting another secret (\ref{sec:attack:secret_extraction}).
As there is no secret accessible to the \ac{amd-sp}'s firmware that we cannot compromise through our attack, updated algorithms are not able to secretly generate, let alone use, keys to restore protection of the non-volatile data.

\subsection{TPM Considerations}

Both users and software and hardware developers can apply various strategies to alleviate the consequences of this attack or regain security for specific applications of a \ac{tpm}.
We will discuss these in the following.

\subsubsection{Using BitLocker protectors effectively}
\label{subsubsec:bitlocker_effectively}

Our attacks have shown that an \ac{ftpm} cannot sufficiently protect its internal state against firmware or physical attacks.
In such a scenario, a \emph{passphrase-only} key protector of reasonable length provides better security than a \emph{TPM-only} protector with a numeric PIN (\ref{sec:case_study:tpm_and_pin:brute_force}).
This is in stark contrast to Microsoft's claim that \enquote{BitLocker provides the most protection when used with a Trusted Platform Module} \cite{microsoft_bitlocker_2021-2} (see also in \ref{sec:background:bitlocker}).
In fact, of all available protectors (seen in Figure \ref{fig:available_protectors}), \emph{TPM-only} is arguably the weakest protection strategy.

However, one can make sense of Microsoft's claim differently:
The \ac{tpm} adds a layer of security when another factor is used.
Specifically, a \emph{TPM and PIN} protector is superior to a \emph{passphrase-only} protection of the same length and character set.
For example, assume an attacker has gained knowledge of the used BitLocker passphrase through social engineering and has physical access to the victim's machine.
The attacker will then be able to boot up Windows and enter the correct PIN to get past the Windows pre-boot prompt.
However, even though the BitLocker-protected volume is now unlocked, the attacker faces the regular Windows login prompt.
At the same time, booting into an attacker-controlled system instead will not enable them to unseal the \ac{vmk}:
Even though they have obtained the correct PIN, changing the boot volume alters the \ac{pcr} registers -- a change the TPM will detect.
In contrast, a \emph{passphrase-only} protector would have enabled him to use, e.g., \emph{Dislocker} to decrypt the protector and mount the volume.

\subsubsection{Easing the setup of `TPM + PIN` protectors}
\label{sec:discussion:tpm_considerations:pin_setup}

BitLocker's \emph{TPM and PIN} protector is currently disabled by default.
Users need to find the respective \emph{Group Policy} settings (\emph{Computer Configuration $\rightarrow$ Administrative Templates $\rightarrow$ Windows Components $\rightarrow$ BitLocker Drive Encryption $\rightarrow$ Operating System Drives $\rightarrow$ Require Additional Authentication at Startup}) before they can use these protectors.
Additionally, they need to use the command-line tool `manage-bde' to set them up \cite{hoffman_how_2017}.
Even worse, Microsoft BitLocker only allows so-called \emph{enhanced PINs} (alphanumeric) with another \emph{Group Policy} change (\emph{Computer Configuration $\rightarrow$ Administrative Templates $\rightarrow$ Windows Components $\rightarrow$ BitLocker Drive Encryption $\rightarrow$ Operating System Drives $\rightarrow$ Allow enhanced PINs for startup}). 

These advanced steps clearly show that Microsoft does not expect or encourage private users to use \emph{TPM and PIN} protectors.
For a non-technical user with high-security requirements, e.g., a journalist, BitLocker's default \emph{TPM-only} configuration gives a false sense of security regarding their encrypted \emph{data-at-rest}.
While \emph{TPM-only} protection relying on \acp{dtpm} is vulnerable to sniffing (\ref{subsubsec:lpc_sniffing_attacks}), our presented attack amends this capability for firmware TPMs.

We believe that it would significantly benefit the security of users if Microsoft provided a way to use \emph{TPM and PIN} in their default BitLocker setup flow.
It would also be thinkable to re-use or derive existing shared user secrets, e.g., the Microsoft Hello PIN, for this matter.
However, \emph{TPM and PIN} protection can be reduced to brute-forcing the PIN when a vulnerable \ac{ftpm} is used (\ref{sec:case_study:tpm_and_pin}).
Since a non-technical user does not discriminate between discrete and firmware TPMs, it would be advisable for Microsoft BitLocker to suggest using an enhanced PIN when an fTPM is used, and implement NIST's password selection rules \cite{burr_electronic_2013}.
Furthermore, we deem it reasonable for Microsoft to distinguish the confusing nomenclature of an \emph{enhanced PIN} from a numeric PIN by introducing a dedicated protector named \emph{TPM and passphrase}, highlighting the improved security level.

\subsubsection{Implementing FDE with \emph{TPM and PIN} securely}
\label{subsubsec:implement_fde_securely}

Our case study shows that \ac{fde} implementations must employ standalone anti-brute-force measures beyond the sealed TPM object as BitLocker does (\ref{sec:case_study:tpm_and_pin_naive}).
If the \ac{tpm} is compromised, this upholds the protector's confidentiality to a degree a (non-TPM) PIN/password-only protector can achieve.
The security of such a method dramatically depends on the length and complexity of the \ac{pin} or password, so strong requirements regarding its length and character set should be considered.

Effective authentication methods are often a trade-off between usability and (cryptographic) strength.
In contrast to \emph{password} and \emph{passphrase}, the term \acf{pin} indicates weaker requirements regarding its length and character set.
For example, credit card \acp{pin} often only consist of 4 numeric digits.
The underlying smart card's lockout mechanism compensates for this low-entropy authentication factor that is potentially prone to brute-forcing attacks.
The \ac{tpm}'s anti-hammering protection pursues a similar goal but is ineffective on \acp{ftpm} compromised by our proposed attack.

\emph{Theoretically}, a layer of encryption could be added to the \ac{tpm} specification:
An object protected by a user authentication policy could be encrypted internally not only by a storage key derived from the parent object but also with a key derived from the user authentication string.
However, this would break basic concepts of the \ac{tpm} specification, e.g., it would no longer be possible to bind together a user authentication policy with another by an "or" clause.
Additionally, the key derivation functions might also prove too expensive for \acp{dtpm}.

\subsubsection{Firmware vs. Discrete TPM}

Consider a TPM application relying exclusively on the TPM to seal a shared secret (like BitLocker's default configuration):
Since our attack is arguably harder to mount than a \ac{dtpm} bus sniffing attack, \acp{ftpm} can be considered more secure than a \ac{dtpm} in this case. 
Apart from this particular case, \acp{dtpm} should be chosen over \ac{ftpm}s for two reasons:
Firstly, since \acp{dtpm} -- to the best of our knowledge -- have not been subject to full state compromises, they protect sensitive data that should never leave the TPM, e.g., private keys.
Secondly, \acp{dtpm} uphold the protection of, e.g., the sealed BitLocker \ac{vmk} protected by \emph{TPM and PIN}, through the anti-hammering protection (see \ref{sec:case_study:tpm_and_pin:brute_force}).
This allows the literal use of low-entropy PINs (compare with Table \ref{tbl:brute-force}).
At the same time, dTPMs make it paramount to use second factors, as they otherwise disclose replayable secrets through their exposed communication interface (as seen with TPM-only BitLocker sniffing attacks).
This significantly limits the practical usage of \ac{dtpm} for server \acf{fde}, as every reboot, e.g., for scheduled security updates, would require human intervention.

\section{Conclusion}
\label{sec:conclusion}

AMD's \aclp{ftpm} of recent microarchitecture generations are vulnerable to physical attacks such as voltage fault injection.
They enable an attacker to access all assets secured by the TPM using low-cost, off-the-shelf hardware.
To the best of our knowledge, our work is the first full TPM state compromise against a widely adopted TPM implementation.
This is a considerable advance compared to previous attacks leveraging external communication to capture replayable secrets or sophisticated side channels compromising select parts of the TPM's internal state.
Our full state compromise gives the powerful ability to defeat any TPM-based security:
Applications relying exclusively on the TPM are left entirely unprotected, while those employing multiple layers of defense face the loss of their TPM-based security layer.

Motivated by Windows 11's push to use the TPM for even more applications,
we apply the vulnerability to Microsoft BitLocker and show the first \emph{fTPM}-based attack against the popular \acl{fde} solution.
BitLocker's default \emph{TPM-only} strategy manages -- without any changes to the user experience -- to swiftly step up a user's security in the face of a lost or stolen device.
However, as our work complements the established attacks against dTPMs with an even more potent attack against AMD fTPMs, a \emph{TPM-only} configuration lulls a non-technical user with high protection needs into a false sense of security.

When attacked with our full state compromise, BitLocker's \emph{TPM and PIN} protector, in contrast, retains a security level according to the PIN's resistance against brute force.
Nonetheless, we find fault that this feature is deeply buried inside Microsoft's \emph{Group Policy} settings and hidden from a non-technical user.
Moreover, a traditional PIN, i.e., a short numeric secret, does not provide even minuscule brute-force protection.
Upgrading the security with so-called \emph{enhanced PINs} -- a euphemism for a \emph{passphrase} -- requires similarly advanced configuration changes.

Microsoft should empower their users to make an informed choice regarding the protection level of their \emph{data-at-rest}: 
Users who fear a physical attacker with reasonable resources should opt for a \emph{TPM and PIN} configuration.
When BitLocker identifies that the underlying TPM is an \ac{ftpm}, users should be urged to turn their \emph{PIN} into a \emph{passphrase}.

While \acp{tpm} are an essential tool to build secure applications, protect and manage cryptographic material, and anchor trust in the hardware of our physical devices, awareness of the required security levels and the \ac{tpm} variant in use is essential.
We hope that our contributions regarding the security of \acp{tpm} in general and AMD's fTPMs in particular guide users and developers on this journey.

\section*{Data Availability}

We publish all code necessary to mount the attack under \cite{jacob_pspreverseftpm_attack_2023}. The repository further includes several intermediate results, e.g., flash memory dumps, to retrace the attack process without possessing the target boards and required hardware tools.

\bibliographystyle{plain}
\bibliography{fTPM-Paper}

\end{document}